\documentclass[a4paper,fleqn,usenatbib]{mnras}

\usepackage[T1]{fontenc}
\usepackage{ae,aecompl}

\usepackage{graphicx}	
\usepackage{amsmath}	
\usepackage{amssymb}	

\usepackage[dvipsnames]{xcolor}

\title[Drivers of stellar metallicity]{SDSS-IV MaNGA: drivers of stellar metallicity in nearby galaxies}

\author[J. Neumann et al.]{
Justus Neumann,$^{1}$\thanks{E-mail: jusneuma.astro@gmail.com}
Daniel Thomas,$^{1,2}$
Claudia Maraston,$^{1}$
Daniel Goddard,$^{1}$\newauthor
Jianhui Lian,$^{3}$
Lewis Hill,$^{1}$
Helena Dom\'inguez S\'anchez,$^{4,5}$
Mariangela Bernardi,$^{4}$\newauthor
Berta Margalef-Bentabol,$^{4}$
Jorge K. Barrera-Ballesteros,$^{6}$
Dmitry Bizyaev,$^{7}$\newauthor
Nicholas F. Boardman,$^{3}$
Niv Drory,$^{8}$
Jos\'e G. Fern\'andez-Trincado,$^{9,10}$
Richard Lane$^{11}$
\\
$^{1}$Institute of Cosmology and Gravitation, University of Portsmouth, Burnaby Road, Portsmouth, PO1 3FX, UK\\
$^{2}$School of Mathematics and Physics, University of Portsmouth, Lion Gate Building, Portsmouth, PO1 3HF, UK\\
$^{3}$Department of Physics \& Astronomy, University of Utah, Salt Lake City, UT 84112, USA\\
$^{4}$Department of Physics and Astronomy, University of Pennsylvania, Philadelphia, PA 19104, USA\\
$^{5}$Institute of Space Sciences (ICE, CSIC), Campus UAB, Carrer de Magrans, E-08193 Barcelona, Spain\\
$^{6}$Instituto de Astronom\'ia, Universidad Nacional Aut\'onoma de M\'exico, A.P. 70-264, 04510 M\'exico, D.F., M\'exico\\
$^{7}$Apache Point Observatory, P.O. Box 59, Sunspot, NM 88349, USA\\
$^{8}$McDonald Observatory, The University of Texas at Austin, 1 University Station, Austin, TX 78712, USA\\
$^{9}$Instituto de Astronom\'ia, Universidad Cat\'olica del Norte, Av.
Angamos 0610, Antofagasta, Chile\\
$^{10}$Instituto de Astronom\'ia y Ciencias Planetarias de Atacama, Universidad de Atacama, Copayapu 485, Copiap\'o, Chile\\
$^{11}$Centro de Investigaci\'on en Astronom\'ia, Universidad Bernardo O'Higgins, Avenida Viel 1497, Santiago, Chile
}

\date{Accepted XXX. Received YYY; in original form ZZZ}

\pubyear{2020}

\begin{document}
\label{firstpage}
\pagerange{\pageref{firstpage}--\pageref{lastpage}}
\maketitle

\begin{abstract}

The distribution of stellar metallicities within and across galaxies is an excellent relic of the chemical evolution across cosmic time. We present a detailed analysis of spatially resolved stellar populations based on $>2.6$ million spatial bins from 7439 nearby galaxies in the SDSS-IV MaNGA survey. To account for accurate inclination corrections, we derive an equation for morphology dependent determination of galaxy inclinations. Our study goes beyond the well-known global mass-metallicity relation and radial metallicity gradients by providing a statistically sound exploration of local relations between stellar metallicity $[Z/H]$, stellar surface mass density $\Sigma_\star$ and galactocentric distance in the global mass-morphology plane. We find a significant resolved mass density-metallicity relation $\rm r\Sigma_\star ZR$ for galaxies of all types and masses above $10^{9.8}\,\mathrm{M_\odot}$. Different radial distances make an important contribution to the spread of the relation. Particularly, in low and intermediate mass galaxies, we find that at fixed $\Sigma_\star$ metallicity increases with radius independently of morphology. For high masses, this radial dependence is only observed in high $\Sigma_\star$ regions of spiral galaxies. This result calls for a driver of metallicity, in addition to $\Sigma_\star$ that promotes chemical enrichment in the outer parts of galaxies more strongly than in the inner parts. We discuss gas accretion, outflows, recycling and radial migration as possible scenarios.


\end{abstract}




\section{Introduction}

Stellar metallicity saves imprints of the star formation activity and recycling efficiency in a galaxy across cosmic time and is key to our understanding of the formation and evolution of galaxies. While gas-phase metallicity reflects the current state of metal abundances and can be understood as the integral over the chemical enrichment (and dilution) history, the metallicity in stars provides evidence of its past evolution.

Numerous works have studied global galaxy parameters in the context of star formation regulation and a set of fundamental relations have been established, in particular between the total stellar mass of a galaxy and its star formation rate \citep[SFR; star formation main sequence, SFMS,][]{Brinchmann2004,Elbaz2007,Noeske2007,Renzini2015}, between the gas density and star formation rate (Schmidt-Kennicutt law, \citealp{Schmidt1959,Kennicutt1998a}) and between the stellar mass and the metallicity, the global mass-metallicity-relation (MZR) both in the interstellar medium \citep{Lequeux1979,Tremonti2004,Kewley2008,Mannucci2010,RosalesOrtega2012,LaraLopez2013,Lian2015,Curti2020} and in stars \citep{Gallazzi2005,Thomas2005,Panter2008,Thomas2010,Kirby2013,Peng2015,Zahid2017,Trussler2020}.

With the advent of integral field spectroscopy (IFS) and the possibility of spatially resolved spectroscopic observations, the study of these fundamental relations have advanced to individual spatial regions of kpc scales within galaxies. There is growing evidence that the global relations have their local counterpart. On kpc scales, stellar surface mass density $\Sigma_\star$ correlates with SFR \citep{Sanchez2013,Wuyts2013,CanoDiaz2016,CanoDiaz2019}, metallicity \citep{RosalesOrtega2012,Sanchez2013,GonzalezDelgado2014,BarreraBallesteros2016,BarreraBallesteros2017} and molecular gas mass \citep{Lin2019b,Lin2020,Ellison2020}.

Spatial variations of galaxy properties have been studied for decades \citep[e.g.][]{Pagel1981, VilaCostas1992} with increasing intensity in recent years owing to the emergence of large observational datasets of IFS surveys, such as SAURON \citep{deZeeuw2002}, ATLAS3D \citep{Cappellari2011}, CALIFA \citep{Sanchez2012}, SAMI \citep{Croom2012,Bryant2015}, MaNGA \citep{Bundy2015} and AMUSING \citep{Galbany2016}. Research has focussed particularly on gradients of metallicity in order to study the assembly history of galaxies as well as inflow and outflow mechanisms.

Most of the works on gas-phase abundance gradients clearly indicate negative slopes with relatively low scatter when the radius is scaled to the size of the galaxy \citep[effective radius $R_\mathrm{e}$ or $R_{25}$;][]{Sanchez2014,Ho2015,SanchezMenguiano2016}. While some studies find a trend of metallicity gradient with stellar mass, in that lower-mass galaxies possess flatter gradients \citep{Belfiore2017,Poetrodjojo2018}, other works suggest no dependence on stellar mass \citep{Lian2018,SanchezMenguiano2018,Bresolin2019}. Other studies have suggested a trend between metallicity gradient and size \citep[e.g.][]{Carton2018}, with \citet{Boardman2021} reporting a significant trend in gradients across the galaxy mass-size plane. The discrepancies seem to arise from different adopted methods to calibrate the abundances, different radial coverage and, potentially, from different spectral resolutions \citep[see e.g.][for a review]{Maiolino2019,Sanchez2020}.

Regarding radial trends of stellar metallicity, there is a general consensus that metallicity gradients are negative in massive galaxies yet become flatter with decreasing mass \citep{Sanchez-Blazquez2014,GonzalezDelgado2015,Goddard2017,Li2018,Lian2018,
DominguezSanchez2019,Oyarzun2019,Lacerna2020}. The correlation with mass seems to be stronger for late-type galaxies (LTGs) than for early type galaxies \citep[ETGs;][]{Goddard2017}.

\defcitealias{GonzalezDelgado2014}{GD14}

During the last ten years, much observational evidence has been provided for a local relation between surface mass density and gas-phase metallicity \citep{RosalesOrtega2012,Sanchez2013,BarreraBallesteros2016} which has also been recently reproduced in cosmological simulations \citep[e.g.][]{Trayford2019}. On the other hand, the spatially resolved \textit{stellar} surface mass density-stellar metallicity relation ($\rm r\Sigma_\star ZR$) has not yet been studied in great detail. It was explored in \citet[][hereafter \citetalias{GonzalezDelgado2014}]{GonzalezDelgado2014} based on a sample of 300 galaxies from the CALIFA survey and it was shortly presented in a new collection of data from different surveys comprising 1494 galaxies in \citet{Sanchez2020}. Interestingly, \citet{Zhuang2019} find a correlation of stellar metallicity with dynamically derived surface mass density from Schwarzschild modeling of 244 CALIFA galaxies. Finally, \citet{Zibetti2020} explored the $\rm r\Sigma_\star ZR$ for 69 CALIFA ETGs. A more in-depth comparison to these works is presented in Sect. \ref{sect:results}.

Given the connection between surface mass density and metallicity, we can partially explain a radially decreasing metallicity with an accordingly decreasing surface brightness profile \citep{deVaucouleurs1959,Sersic1968,Freeman1970} and, thus, a decreasing surface mass density \citep{Bakos2008,GarciaBenito2019}. However, it is not yet clear whether the surface mass density is sufficient in explaining radial variations of stellar metallicity or whether other additional drivers of metallicity may play a role, such as the total stellar mass, the morphological type, the environment, accretion and outflow of gas, or structural galaxy components such as galaxy bars and bulges.

In the present work, we aim to investigate the distribution of stellar metallicity in nearby galaxies as a function of galactocentric radius and stellar surface mass density in the global mass-morphology plane. We make use of an unprecedented large sample of galaxies from the MaNGA survey, which allows us to explore the relations in several bins of stellar mass and morphology while keeping the statistics sufficiently large.

The paper is organised as follows. In Sect. \ref{sect:data}, we give a brief outline of the MaNGA survey and our sample selection. This is followed by a presentation of our analysis in Sect. \ref{sect:analysis}, where we introduce an update to our MaNGA Firefly catalogue and derive an empirical prescription for galaxy inclination corrections. Thereafter, we present our results in Sect. \ref{sect:results}, discuss our findings in Sect. \ref{sect:discuss} and conclude with Sect. \ref{sect:conclusion}.

Throughout the paper we assume a flat cosmology with a Hubble constant of $H_0 = 67.8\mathrm{\,km\,s^{-1}\,Mpc^{-1}}$ and $\Omega_\mathrm{m} = 0.308$ \citep{Planck2016}.




\section{Data}
\label{sect:data}

\subsection{The MaNGA survey}

This work is part of the Mapping Nearby Galaxies at Apache Point Observatory survey \citep[MaNGA,][]{Bundy2015}, which is a Sloan Digital Sky Survey-IV project \citep[SDSS-IV,][]{Blanton2017}. MaNGA uses specially designed integral field unit fibre systems \citep{Drory2015} that feed into the BOSS Spectrograph \citep{Smee2013} mounted at the Sloan Foundation 2.5-meter Telescope \citep{Gunn2006}.

Each observing plate employs an array of 17 hexagonal bundles of different sizes between 19 and 127 fibres to ensure a uniform spatial coverage per galaxy out to at least $\rm 1.5\,R_e$ for the Primary+ MaNGA sample ($\sim$2/3 of the total sample) and out to $\rm 2.5\,R_e$ for the higher redshift Secondary sample \citep[$\sim$1/3;][]{Law2015,Wake2017} at a median spatial resolution of $2.54\,\arcsec$ FWHM \citep{Law2016}. Objects are observed 2-3$\rm \,hr$ to reach the signal-to-noise (S/N) per pixel required by the survey goals \citep{Yan2016b}. The spectra cover a wavelength range of 3622$\,$\AA-10354$\,$\AA$\ $at a spectral resolution of $\rm \sigma = 72\,km\,s^{-1}$ \citep{Law2016}.

Raw observations are spectrophotometrically calibrated \citep{Yan2016a} and reduced by the Data Reduction Pipeline \citep[\texttt{DRP},][]{Law2016}. High-level data products for each galaxy such as stellar kinematics, emission line properties and spectral indices are produced by the Data Analysis Pipeline \citep[\texttt{DAP},][]{Belfiore2019, Westfall2019}.

The MaNGA survey has concluded its observations in August 2020. The final sample
encompasses $10,010$ unique galaxies across $\sim 4,000\,\mathrm{deg^2}$ at a median redshift of $z\sim 0.03$ and will be released by the end of 2021. The sample was selected to have a flat stellar mass distribution in logarithmic space in the range $5\times 10^8\,M_{\sun} \leq M_{\star} \leq 3\times 10^{11}\,M_{\sun}$. Additionally, a colour-enhanced supplement sample was added to upweight underrepresented galaxies in the colour-magnitude diagram. The full main MaNGA sample consists of the Primary sample ($\sim 50\%$) and its Colour-Enhanced supplement ($\sim 17\%$; together called Primary+), with radial coverage out to $\rm 1.5\,R_e$, and the Secondary sample ($\sim 33\%$), with radial coverage out to $\rm 2.5\,R_e$ \citep[for more details, see][]{Wake2017}.

\subsection{Sample selection}

The parent sample of the present work is drawn from the final MaNGA Product Launch 11 (MPL-11), which will be released in the upcoming 17th SDSS data release (DR17). The data comprises 11273 datacubes, 10145 of which represent high-quality galaxy observations coresponding to 10010 unique galaxies and 135 repeat observations. From the 10010 galaxies, we selected all those that are part of the main MaNGA sample\footnote{There is an overlap between the main sample and the MaNGA ancillary programmes. This first selection only requires the galaxy to be part of the main sample and it does not deselect galaxies that are also part of an ancillary programme.} and that have been analysed by the \texttt{DRP} and \texttt{DAP}, which yields a sample of 9714 galaxies.


From this sample, we removed 
\begin{itemize}
\item datacubes that were classified as critical by the \texttt{DRP} or by the \texttt{DAP},
\item galaxies from the galaxy pair ancillary programmes,
\item galaxies with inclinations $i>80\degr$ (see Sect. \ref{sect:incl}),
\item datacubes with less than 30 Voronoi bins for a given galaxy (to assure decent spatial sampling, see Sect. \ref{sect:FireflyVAC}).
\end{itemize}


Our analysis further requires a morphological classification. For that purpose we cross-matched our sample with that of the MaNGA Deep Learning Morphological Value-Added-Catalogue \citep[MDLM-VAC-DR15;][]{Fischer2019} that uses Deep Learning algorithms based on Convolutional Neural Networks described in more detail in \citet{DominguezSanchez2018}. In this work, we use an updated version of the MDLM-VAC that contains 10293 unique entries from the upcoming DR17, kindly provided by H. Dom\'inguez S\'anchez and collaborators (MDLM-VAC-DR17; Dom\'inguez S\'anchez et al. in prep.). Morphological $T$-types are obtained from models that have been trained and tested with the morphological galaxy catalogue from \citet{Nair2010} on RGB SDSS-DR7 cutout images. Values of the $T$-types range from $-4$ to $9$. Two additional classifications from the catalogue are used in this paper: $P_{S0}$ for a finer separation between ellipticals and lenticulars, and $P_{LTG}$ to help separating ETGs from LTGs in addition to the $T$-types. The value given in the catalogue is the average of the output of ten different models trained with k-folding. Finally, we also include the visual classification flag $VC$. We use the following definition to separate between ellipticals ($T \leq 0$ \& $P_{LTG}<0.5$ \& $P_{S0}<0.5$ \& $VC=1$), lenticulars ($T \leq 0$ \& $P_{LTG}<0.5$ \& $P_{S0}\geq0.5$ \& $VC=2$) and LTGs ($T>0$ \& $P_{LTG}\geq 0.5$ \& $VC=3$). The cross-match of our selected sample with the objects in the MDLM-VAC that have a morphological classification with the aforementioned criteria yields our final working sample of 7439 galaxies (2215 Es, 808 S0s, 4416 LTGs).

\begin{figure}

	\includegraphics[width=\columnwidth]{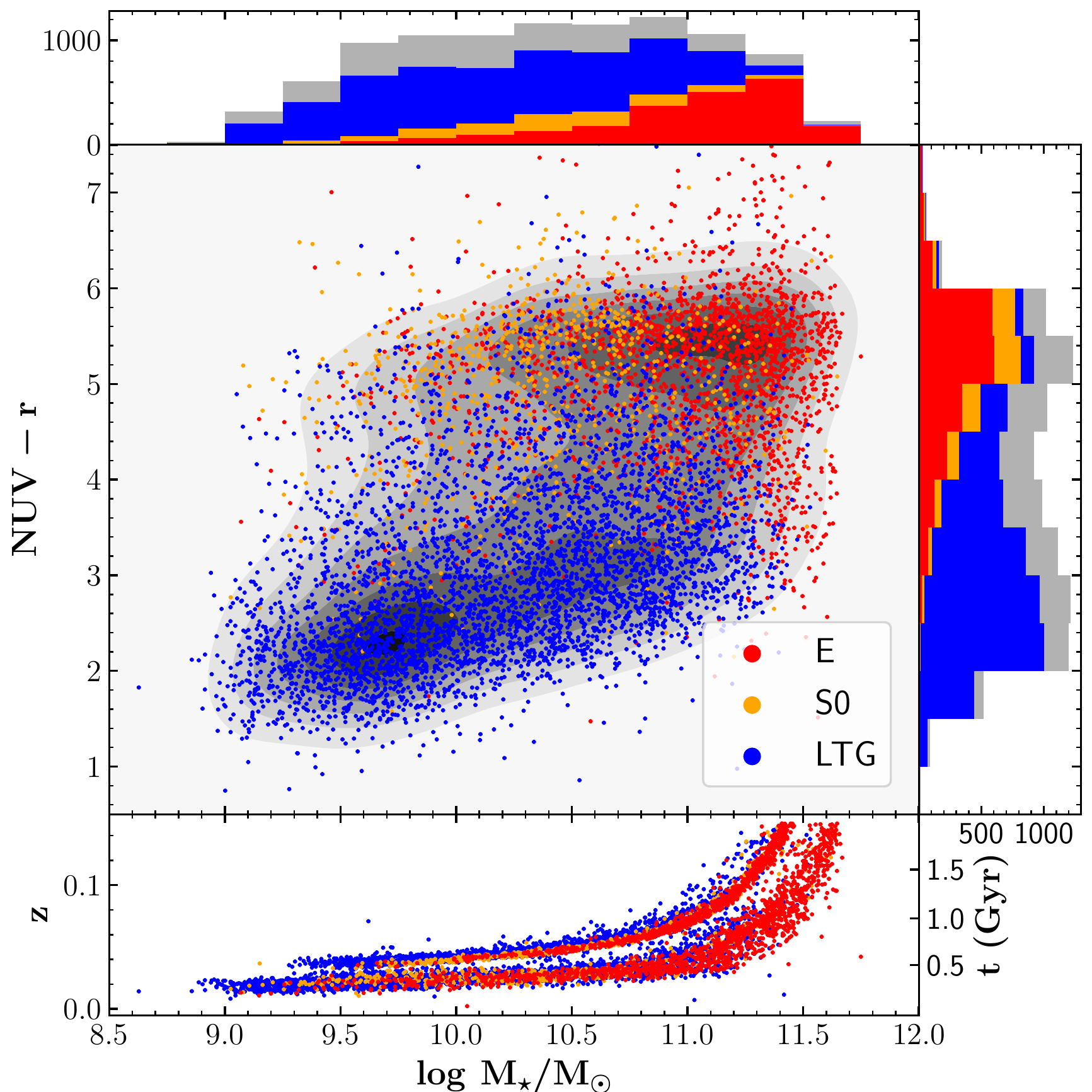}
    \caption{\textit{Top:} Colour-mass diagram derived from SDSS photometric masses, $NUV$- and $r$-band filters, taken from the NSA catalogue. Grey contours and histograms show the complete MaNGA DR17 main sample. Scatter points show individual galaxies from the selected working sample separated by morphology into ellipticals (E, red), lenticulars (S0, orange) and late-type galaxies (LTG, blue). Histograms for E, S0 and LTG are stacked. \textit{Bottom:} Redshift-mass distribution of the sample. Redshifts are taken from the spectroscopical estimate `STELLAR\_Z' in the \texttt{DAP}.}
    \label{fig:sample_colour}
\end{figure}

Figure \ref{fig:sample_colour} presents the sample in the $NUV - r$ colour-mass and redshift-mass diagram in comparison with the main DR17 sample. We use total galaxy masses from the NASA-Sloan Atlas (NSA)\footnote{M. Blanton; \url{www.nsatlas.org.}} that were obtained from K-correction fits to elliptical Petrosian photometric fluxes \citep{Blanton2011,Blanton2017a}, assuming a \citet{Chabrier2003} initial mass function (IMF) and \citet{Bruzual2003} single stellar population models (SSPs). These fits are held to contain nearly all flux of a galaxy and do not depend on aperture correction. We use the total galaxy masses in this work for the global relations shown in Sect. \ref{sect:global} and to separate the sample in stellar mass bins. We scaled the masses to the Hubble parameter $H_0 = 67.8\mathrm{\,km\,s^{-1}\,Mpc^{-1}}$ used in this work.

There is no obvious bias in the sample and our selection seems representative of the complete MaNGA sample. The distribution is close to flat over a large range of mass and color. Galaxies classified in the MDLM-VAC as LTGs occupy predominantly the blue cloud whereas S0s and Es are located on the red sequence and dominate the high-mass end of the distribution. The gap seen in the redshift distribution is a natural consequence of the higher redshift selection criterion of the Secondary sample as compared to the Primary+ sample.

\section{Analysis}
\label{sect:analysis}

The results presented in this work are based on measurements of spatially resolved stellar population properties, such as stellar metallicities ($[Z/H]$), stellar ages, masses ($M_\star$) and surface mass densities ($\Sigma_\star$). The derivation of these parameters is a multi-step process that is outlined in the following.

\subsection{MaNGA Firefly VAC}
\label{sect:FireflyVAC}

The extraction of the stellar population properties is done by employing the full-spectral-fitting technique using our own \texttt{Firefly}\footnote{\url{https://www.icg.port.ac.uk/firefly/}} code \citep{Wilkinson2017}. \texttt{Firefly} is a $\chi^2$ minimisation code that fits combinations of single-burst stellar population models (SSPs) to spectroscopic data.

The \texttt{Firefly} analysis of MaNGA data builds upon the \texttt{DAP}, which already includes all necessary preparatory steps. In particular, the \texttt{DAP} workflow include (1) adaptive Voronoi spatial-binning \citep{Cappellari2003} to a minimum target $\rm S/N \sim 10$ in the stellar continuum, (2) measurements of stellar kinematics and (3) emission line fluxes employing the full-spectral-fitting code \texttt{pPXF} \citep{Cappellari2004,Cappellari2017}. The \texttt{DAP} uses a hierarchically clustered (HC) set of template spectra from the MILES stellar library \citep{Sanchez-Blazquez2006} for the extraction of stellar kinematics, while the stellar continuum modelling in the emission line module uses a subset of MaStar SSPs \citep{Maraston2020} based on the MaStar stellar library \citep{Yan2019}.

From each \texttt{DAP}-processed datacube, \texttt{Firefly} reads the Voronoi-binned spectra, subtracts emission line fluxes and shifts the spectra to restframe wavelengths using the redshift and stellar velocity information from the \texttt{DAP}. Additionally, the stellar velocity dispersion and the instrumental resolution are used to match the line broadening of the SSPs to the data prior to the fitting process. For the results presented in this work, we used the stellar population models of \citep{Maraston2011} based on the MILES stellar library (m11-MILES) and a \citet{Kroupa2001} IMF. Updates and comparison using our new MaStar models \citep{Maraston2020} are subject for future papers. 

\texttt{Firefly} fits arbitrarily weighted linear combinations of SSPs to the data following an iterative process controlled by the Bayesian Information Criterion. No additive or multiplicative polynomials are employed and star formation histories (SFHs) are not regularised in order to allow for a large fitting freedom and sufficient exploration of the parameter space. \texttt{Firefly} has been shown to perform well down to $\rm S/N \sim 5$ \citep[see][for more detail]{Wilkinson2017}.

The results of the complete \texttt{Firefly} run on all DR17 datacubes are stored in the MaNGA Firefly Value-Added-Catalogue (VAC; Neumann et al. in prep.)\footnote{Note that the DR17 version of this catalogue has not yet officially been released. See \url{https://www.sdss.org/dr15/manga/manga-data/manga-firefly-value-added-catalog/} for the latest public release of the MaNGA Firefly VAC. A final release is planed for the upcoming SDSS DR17 by the end of 2021.}. This constitutes an update of the earlier versions of the VAC \citep{Goddard2017,Parikh2018} and  provides spatially resolved stellar population properties, including stellar mass and its partition into stellar remnants, age, metallicity, stellar surface mass density, dust attenuation and the full SFHs for 10735 datacubes. The MaNGA Firefly VAC complements the \texttt{DAP} by providing further high-level data products in addition to the stellar kinematics, emission line properties and spectral indices given by the \texttt{DAP}.

Throughout the analysis in this paper, we use light-weighted and linearly averaged stellar population parameters. In most cases, however, we employ the median, which is identical in linear and logarithmic space. In addition to our galaxy sample selection, we apply a $\rm S/N>8$ cut to all Voronoi bins. Furthermore, we correct all stellar ages to the median redshift of the sample $z=0.0376$.

\subsection{Inclination correction}
\label{sect:incl}

\begin{figure}
	\includegraphics[width=\columnwidth]{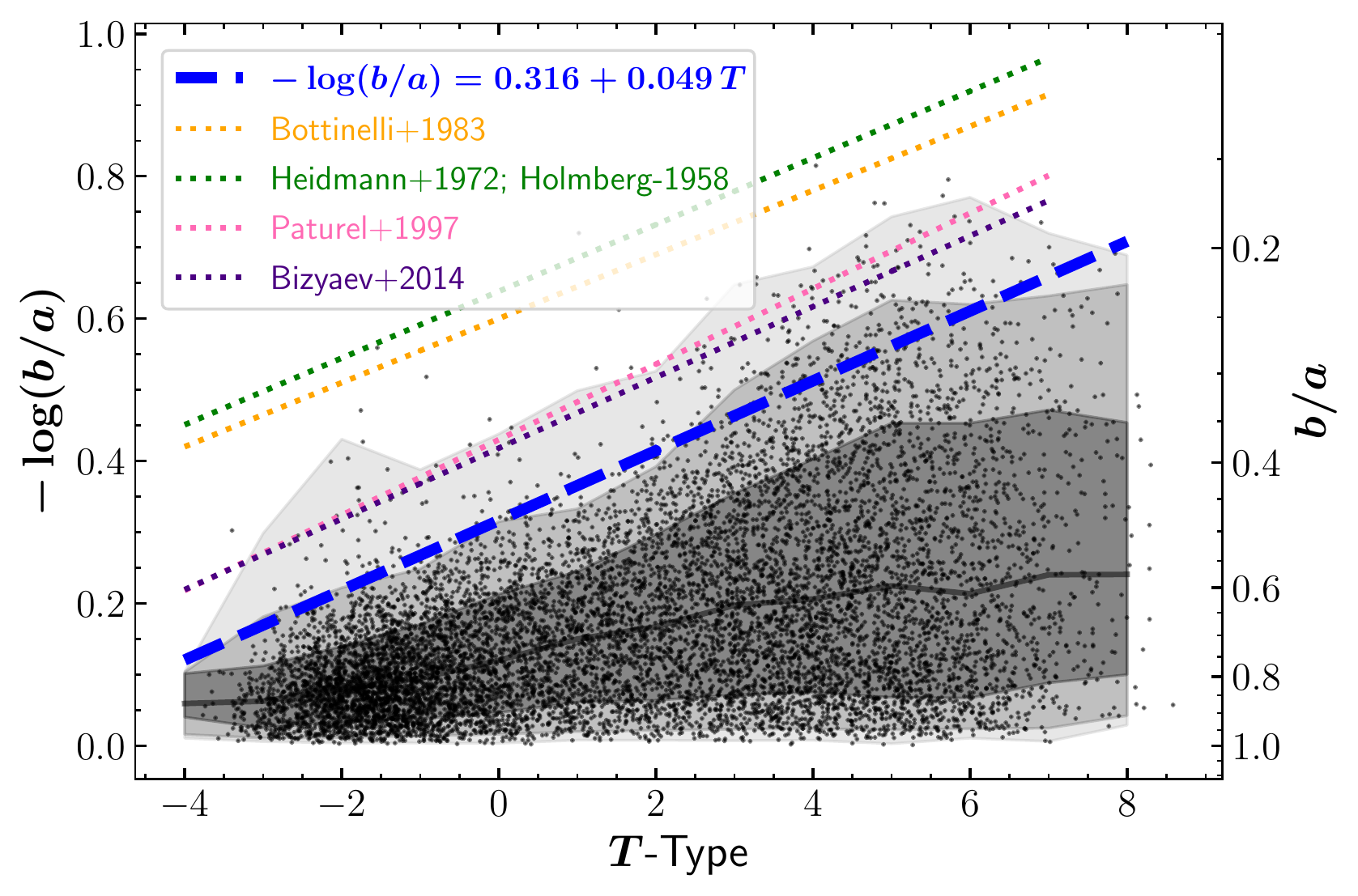}
    \caption{Axial ratios $ q = b/a$ as a function of morphological $T$-type. Black points are individual galaxies. The dark grey line shows the running median and the grey shaded areas from dark to light mark the inner $68\%$, $95\%$ and $99\%$ around the median, respectively. The blue dashed line is a linear fit to the 2.5th percentile.}
    \label{fig:incl}
\end{figure}

\begin{figure}
	\includegraphics[width=\columnwidth]{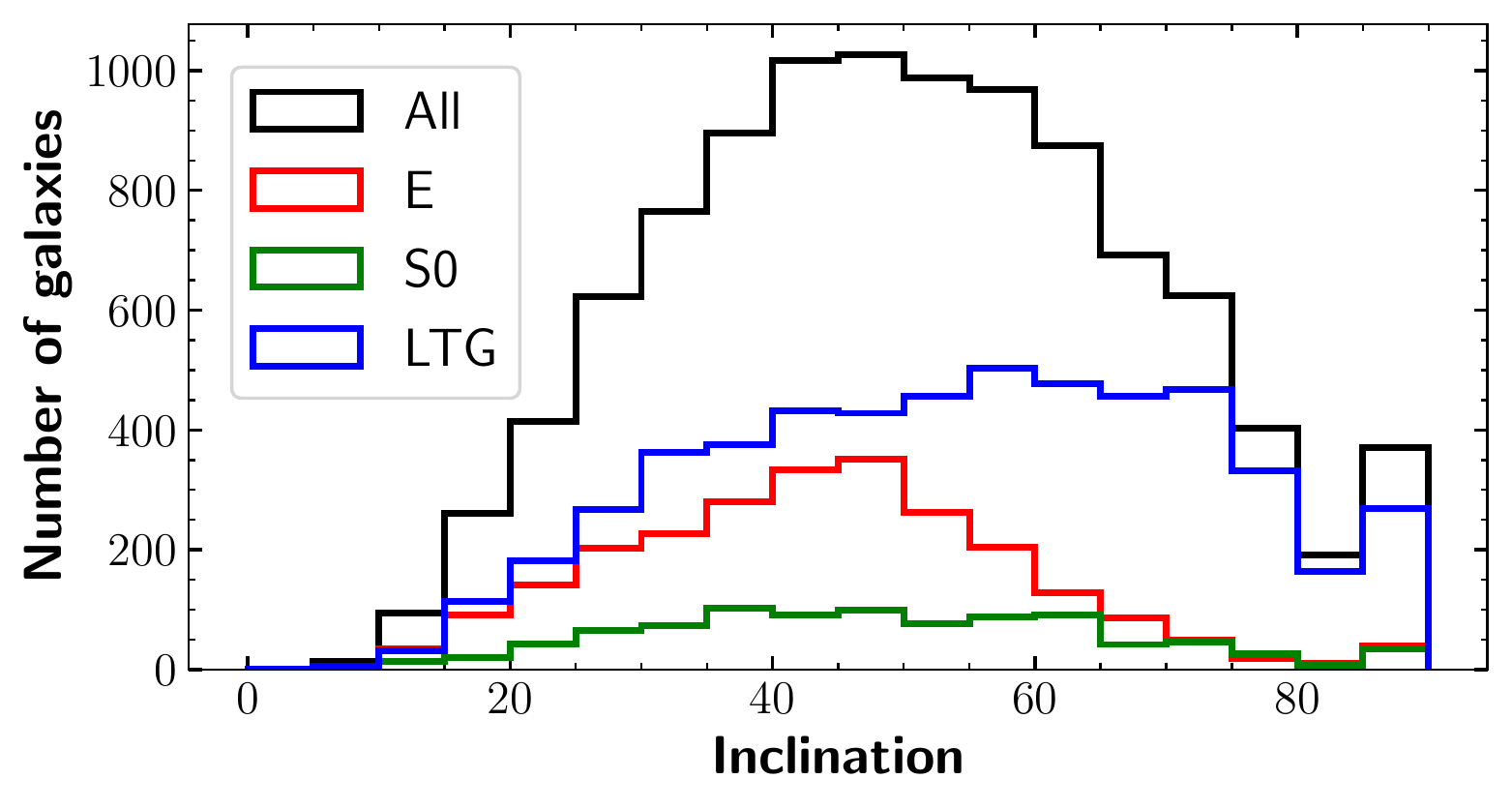}
    \caption{Distribution of galaxy inclinations derived with equation \ref{eq:q0} from Fig. \ref{fig:incl}.}
    \label{fig:incl_dist}
\end{figure}

The stellar surface mass density $\Sigma_\star$ is defined as the stellar mass divided by the surface area. The stellar masses are calculated by \texttt{Firefly} for each spatial Voronoi bin and $\Sigma_\star$ is given in the MaNGA Firefly VAC, but the area of the corresponding bin still needs to be corrected for projection effects. For a circular thin disc the deprojected area is given by $A = A_\mathrm{obs} \cdot (\cos i)^{-1}$, where $A_\mathrm{obs}$ is the observed area and $i$ is the inclination of the disc. This yields $\Sigma_\star = \Sigma_{\star,\,\mathrm{obs}} \cdot \cos i$ for the surface mass density. However, real galaxies have an intrinsic thickness $q_0$, the exact value of which is often very uncertain and varies with morphology \citep{Heidmann1972,Bottinelli1983,Guthrie1992}, wavelength \citep{Mitronova2004} and kinematics \citep{Weijmans2014}. The inclination of a thick oblate spheroid is obtained as $\cos i = \sqrt{(q^2-q_0^2)/(1-q_0^2)}$, where $q$ is the observed axial ratio of the projected spheroid \citep{Hubble1926}. Thus, we have

\begin{equation}
\Sigma_\star = \Sigma_{\star,\,\mathrm{obs}} \cdot \sqrt{(q^2-q_0^2)/(1-q_0^2)}.
\label{eq:sigma}
\end{equation}

The apparent axial ratios of the galaxies in our sample are obtained from the elliptical Petrosian analysis in the enhanced NSA catalogue \citep{Wake2017}.

Under the simplification that the intrinsic axial ratios only vary with morphology and that galaxies seen face-on are perfectly circular \citep[but see e.g.][who find that discs are slightly elliptical]{Ryden2004}, we derive the intrinsic thickness of each galaxy following a similar procedure as described in \citet{Heidmann1972}, \citet{Bottinelli1983} and \citet{Guthrie1992}.

In Fig. \ref{fig:incl}, we plot the observed axis ratio of each galaxy in the sample against the morphological $T$-type. In this approach, the assumption is that at a given $T$-type the galaxies with the lowest axis ratio are seen edge-on and, thus, for these galaxies $q_0 = q$. We fit a linear function in log-linear space to the lower 2.5th percentile of the sample to allow for scatter due to measurement errors, as seen in the figure. This yields the relation

\begin{equation}
-\log q_0 = 0.316 + 0.049\,T.
\label{eq:q0}
\end{equation}

We tested whether the derivation of $q_0$ is dependent on galaxy mass by selecting galaxies of different mass bins and reevaluating Fig. \ref{fig:incl} and found no significant change. This equation is directly empirically derived from the same sample of MaNGA galaxies for that we aim to derive the inclinations in this work, but we would like to emphasise that it can be used for any kind of galaxy given the assumptions stated above. 

In comparison with our analysis, we also show in Fig. \ref{fig:incl} the relations derived by \citet[][using the measurements from \citealt{Holmberg1958}]{Heidmann1972}, \citet{Bottinelli1983}, \citet{Paturel1997} and \citet{Bizyaev2014}. The slope of our equation (0.049) is very similar to those in the literature (0.047, 0.045, 0.053, and 0.0497, respectively). However, the derivation of the intercept is clearly different, with our measurements yielding the largest axis ratios of edge-on systems. Part of the disagreement stems from the somewhat arbitrary choice of allowing for a certain amount of scatter above the fitted upper limit. The more recent results from the mentioned literature, \citet{Paturel1997} and \citet{Bizyaev2014}, for example, agree well with our statistics, would we be more stringent with the scatter. In addition to that, the measurement of the axis ratios of galaxies is sensitive to the applied technique. For instance, the authors cited above use ellipse fitting at a surface brightness level of 25 and 26.5$\rm \,mag\,arcsec^{-2}$ or at a certain S/N level, while the axis ratios from the NSA catalogue employed here are derived at the 90\% Petrosian light radius. Furthermore, some authors who find very low $q_0 \leq 0.2$ for LTGs correct axial ratios for the distortions of the bulge component \citep[cf.][]{Giovanelli1994} what naturally leads to smaller ratios for smaller $T$-types.

Comparing our results with the kinematic analysis of ETGs in \citet{Weijmans2014}, we find a very good agreement for our galaxies with $T \leq 0$ and $0.75>q_0>0.48$ and their results for slow rotators: $q_0 = 0.63 \pm 0.09$. Fast rotators, on the other hand, have been found to be much flatter with $q_0 = 0.25 \pm 0.14$, a range of axis ratios that is only populated by LTGs in Fig. \ref{fig:incl}.

Using equations \ref{eq:sigma} and \ref{eq:q0}, we are now able to determine the inclination and intrinsic surface mass density. The distribution of inclinations is shown in Fig. \ref{fig:incl_dist}. To further test our results we compared a subset of $\sim 450$ galaxies to the kinematically derived inclinations for the same galaxies in Yang et al. (submitted) and found a good agreement.

In addition to conducting an inclination correction to the surface mass density, we decided to exclude galaxies with $i > 80 \degr$ from the further analysis, because edge-on galaxies are prone to large uncertainties in the derivation of integrated stellar population measurements. We tested how different inclination cuts affected our analysis and found that the main results of this work remain unchanged, but including galaxies with $i > 80 \degr$ introduced increased scatter into the derived relations.



\section{Results}
\label{sect:results}

In this section, we present the results of our analysis. We start by reproducing the well-known global mass-metallicity relation (MZR), but separated by morphology, and discuss it in light of previous studies. Afterwards, we zoom-in on spatially resolved relations starting with radial variations of stellar metallicity before presenting the complex interplay between metallicity, surface mass density, galactocentric distance, total mass and morphology from different angles.

\subsection{Global stellar mass-age and mass-metallicity relation}
\label{sect:global}

The global MZR is a relation between the total stellar mass of a galaxy and its metallicity. For this relation, measurements of the central metallicity \citep[e.g.][]{Trussler2020}, the metallicity at the effective radius $R_\mathrm{e}$, the galaxy-wide averaged metallicity \citepalias{GonzalezDelgado2014} or the mean metallicity within $R_\mathrm{e}$ \citep{Sanchez2018} have been used in the literature. In all cases a clear correlation has been found, indicative of an underlying local relation.

\begin{figure}
	\includegraphics[width=\columnwidth]{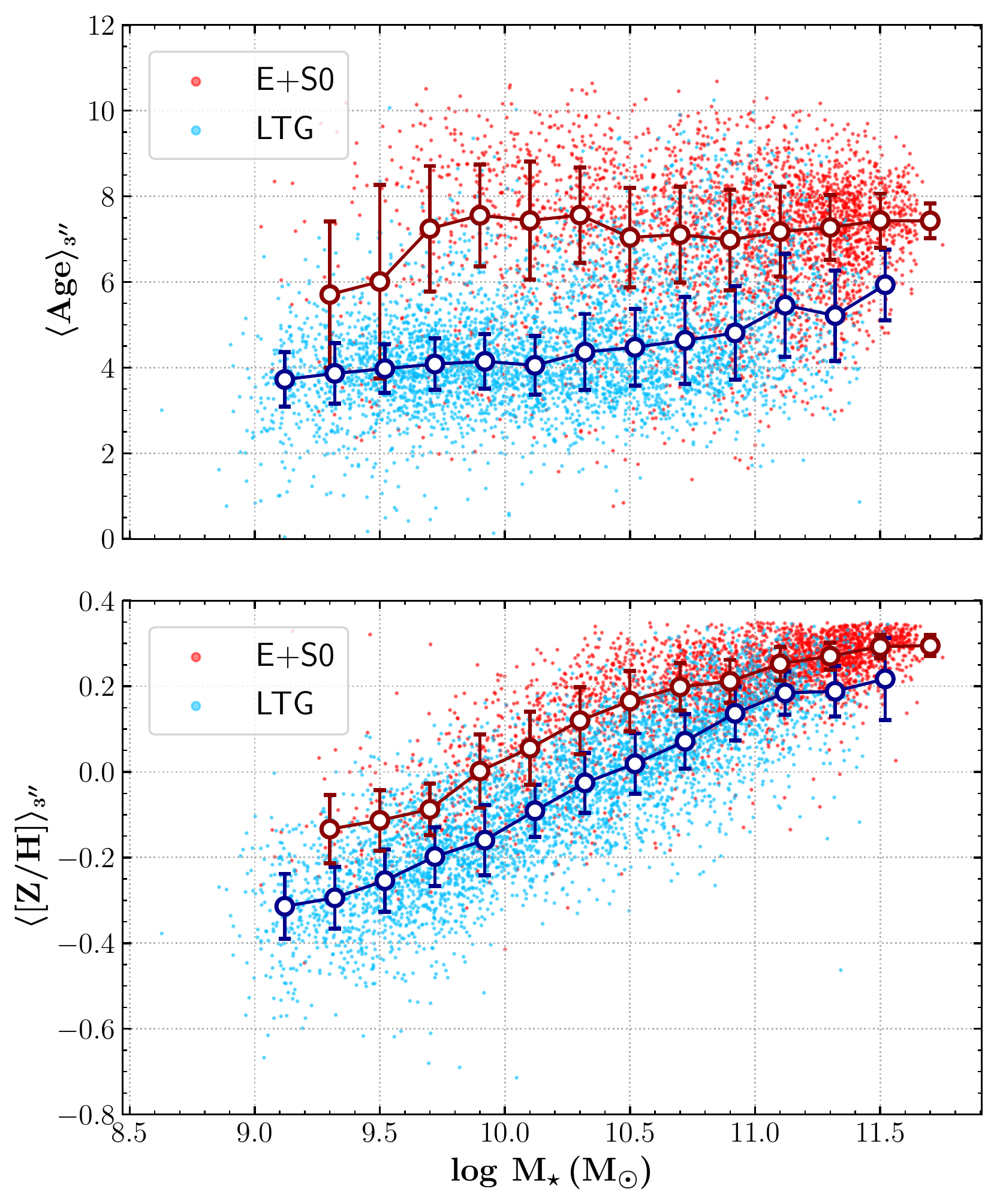}
    \caption{Global mass-age relation (top) and mass-metallicity relation (bottom) separated by morphology. Age and metallicity are averaged within a central area of $3\arcsec$ diameter. The scatter points represent individual galaxies and the line plot shows the median age and metallicity across mass bins of $0.2\,$dex width. The error bars display the median absolute deviation.}
    \label{fig:MZR_glob}
\end{figure}

In Fig. \ref{fig:MZR_glob}, we show the global mass-age and mass-metallicity relation of ETGs and LTGs in our sample. Both age and metallicity are averaged within the central $3\arcsec$ diameter for a better comparison with previous SDSS-based literature. These values are provided in our MaNGA Firefly VAC. We do not find a significant difference regarding the central stellar population properties between elliptical and lenticular galaxies and decided to combine them here as ETGs. They will be shown separately in the rest of the work.

The global mass-age relation for LTGs is mildly positive for $M_\star<10^{10}\,\mathrm{M_\odot}$ with a stronger increase towards higher masses. The ages of ETGs are systematically older and the distribution across different masses is also relatively flat with a slightly positive age-mass correlation. These trends are in good agreement with results from \citet{Trussler2020} who studied the global chemical properties of $\sim$ 80,000 local galaxies in SDSS, separated by SFR into passive, star-forming and green valley galaxies. In this comparison we assume that the morphological classification used in our work is a good first-degree approximation of star-forming and passive galaxies in \citet{Trussler2020}. It is important to point out that while the trends are similar, the absolute values differ owing to the fact that we show light-weighted quantities, while they show mass-weighted values, the latter of which are usually older and more metal-poor.

We observe a clearly positive mass-metallicity relation of both ETGs and LTGs with very similar slopes, but ETGs are systematically offset towards higher metallicities. This is again in agreement with \citet{Trussler2020} and has been explained by the authors as due to a combination of an extended starvation phase and outflows \citep[see also][]{Peng2015}. In short, the star-forming progenitor of a passive galaxies stops accreting new gas during the quenching phase and consequently uses up the remaining gas in the ISM efficiently and without dilution to form new metal-rich stars, which leads to a sharp increase in stellar metallicity while stellar mass increases only slowly.

In summary, we find a mildly positive global mass-age relation and a positive mass-metallicity relation, with the central parts of ETGs being systematically older and more metal-rich than LTGs at fixed total mass. The positive MZR from our IFS data reproduces well earlier findings from both single-fibre observations \citep[e.g.][]{Thomas2010,Zahid2017,Trussler2020}, semi-analytic models \citep{Henriques2020,Yates2021} and cosmological simulations \citep[e.g.][]{Schaye2015,Ma2016,deRossi2017}.

\begin{figure*}
	\centering
	\includegraphics[width=17cm]{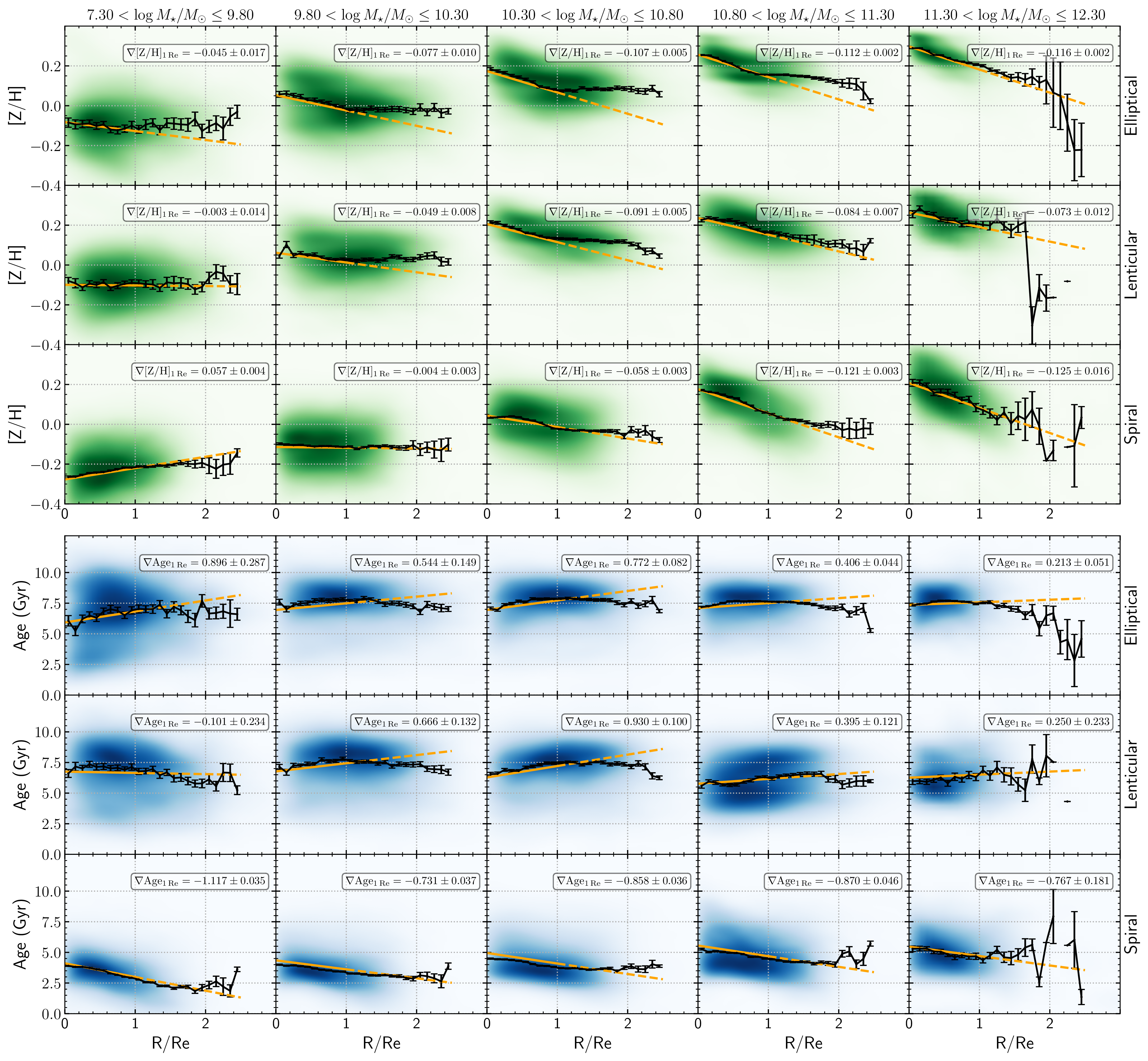}
    \caption{Radial metallicity (top) and age (bottom) profiles. Columns show different bins of total stellar mass $M_\star$. Rows show different morphologies. Each panel presents a density plot of individual spatial bins from galaxies with the corresponding mass and morphology. The data are smoothed with a Gaussian kernel density estimator. In black, we show median $[Z/H]$ and age in bins of $0.1\,R_\mathrm{e}$ width. The error bar is the standard error on the median, calculated as $\sigma_\mathrm{err}=(\pi/2)\cdot (\sigma/\sqrt{N})$, where $\sigma$ is the standard deviation and $N$ the sample size per bin. A linear regression to the data within $1\,R_\mathrm{e}$ is plotted as solid orange line with an extrapolation towards larger radii as dashed line. The corresponding slope is annotated in each panel.}
    \label{fig:ZR_and_AR}
\end{figure*}

\subsection{Radial trends of stellar populations}

After testing the global galaxy-to-galaxy variations of stellar populations with stellar mass and morphology, we are now in the position to further exploit the potential of the IFS data and study stellar population parameters locally as a function of galactocentric distance. Henceforth, we separate our galaxy sample in three morphology bins (elliptical, lenticular, spiral) times five total stellar mass bins ($\log M_\star/M_\odot \in [7.3,9.8,10.3,10.8,11.3,12.3]$). We further average the spatially resolved parameters (from the Voronoi bins) of each galaxy individually in radial bins of $0.05 \times R_\mathrm{e}$ size. This is done to prevent galaxies with many Voronoi bins from being upweighted and, thus, in order to assure that each galaxy contributes equally to the subsequent analysis. In the following, when we talk about individual spatial bins, we are not referring to the Voronoi bins but to the radial bins. Examples of 2D stellar population maps produced by \texttt{Firefly} from nine galaxies across the mass-morphology grid can be found in Fig. \ref{fig:galEx}.

\subsubsection{Age and metallicity profiles}

In Fig. \ref{fig:ZR_and_AR}, we present radial age and metallicity profiles of our sample. The coloured density plots in the background show the distribution of individual spatial bins, while the black line shows the median in radial bins of $0.1\,R_\mathrm{e}$. Additionally, we show a linear fit to the profile within $1\,R_\mathrm{e}$, which is plotted as solid orange line  for $R \leq R_\mathrm{e}$ and as dashed extrapolated line for larger radii.

We observe negative inner metallicity gradients for all galaxies with $\log M_\star/M_\odot>9.8$, which flatten towards larger radii, most clearly seen for Es and intermediate mass S0s. For the lowest-mass galaxies, metallicity profiles of Es and S0s are very shallow across the whole radial range and are clearly positive ($\nabla[Z/H]_{1\,R\mathrm{e}} = 0.057 \pm 0.004\,\mathrm{dex/R_e}$) for spiral galaxies. Furthermore, it is interesting to note that while the metallicity changes dramatically in the centre across the mass-morphology plane -- giving rise to the global MZR as presented in the previous section -- the outermost parts at $R=2.5\,R_\mathrm{e}$ do not change much. This could be an indication of a relatively uniform stellar metallicity in the outer halo of galaxies. However, we do not sample these regions well enough to draw firm conclusions.

The age profiles show relatively flat, mildly positive gradients for elliptical and lenticular galaxies for all masses ($|\nabla \mathrm{Age_{1\,Re}}|<1\,\mathrm{Gyr/R_e}$). They are negative but shallow for spiral galaxies with a central steepening at high masses ($\log M_\star/M_\odot>10.8$), indicative of an old bulge component. Nevertheless, it is apparent in the density plot that not all LTGs in the high-mass bin have this central upturn. In fact, a substantial fraction seems to have a rather flat age distribution pointing towards the absence of a dominating classical bulge. Lenticular galaxies show an interesting transition in the age distribution from mostly older ages for $\log M_\star/M_\odot<11.3$ to younger ages for $\log M_\star/M_\odot>10.8$, with a bimodality in between which is typically observed between active and passive galaxy populations. This could be related to the differences seen in the stellar population gradients between low-mass and high-mass S0s in \citet{DominguezSanchez2020}, but needs yet to be investigated in more detail.

Increasing the sample size of previous works by more than an order of magnitude, we find that the median age and metallicity gradients presented here in light-weighted quantities and their trends with mass and morphology are qualitatively in good agreement with previous analysis of MaNGA data with the \texttt{Firefly} full spectral fitting code \citep{Goddard2017} and other literature \citep[e.g.][]{Kuntschner2010,Sanchez-Blazquez2014,GonzalezDelgado2015,Li2018,Lian2018,
Oyarzun2019,Zhuang2019,Lacerna2020,Sanchez2020}. However, the metallicity gradients for massive ellipticals ($\nabla[Z/H]_{1\,R\mathrm{e}} = -0.116 \pm 0.002\,\mathrm{dex/R_e}$) and lenticulars ($\nabla[Z/H]_{1\,R\mathrm{e}} = -0.073 \pm 0.012\,\mathrm{dex/R_e}$) are flatter than some results from the literature indicate \citep[e.g.][]{Kuntschner2010,MartinNavarro2018,DominguezSanchez2019,Zibetti2020}, yet they are consistent with other studies \citep[e.g.][]{Mehlert2003,GonzalezDelgado2015,Goddard2017,Li2018,Lacerna2020}. Some of the discrepancy can be attributed to differences in stellar population analysis techniques (full spectral fitting, line strength measurement), averaging (light-weighted, mass-weighted, linear, logarithmic), sample selection and data quality (S/N, sampling), see e.g. \citet{Goddard2017} and \citet{Zibetti2020} for further discussion. The detailed variations seen in the density distributions of single spatial bins across the mass-morphology plane leaves room for further research in upcoming papers.

\subsubsection{Surface mass density profiles}

\begin{figure}
	\centering
	\includegraphics[width=\columnwidth]{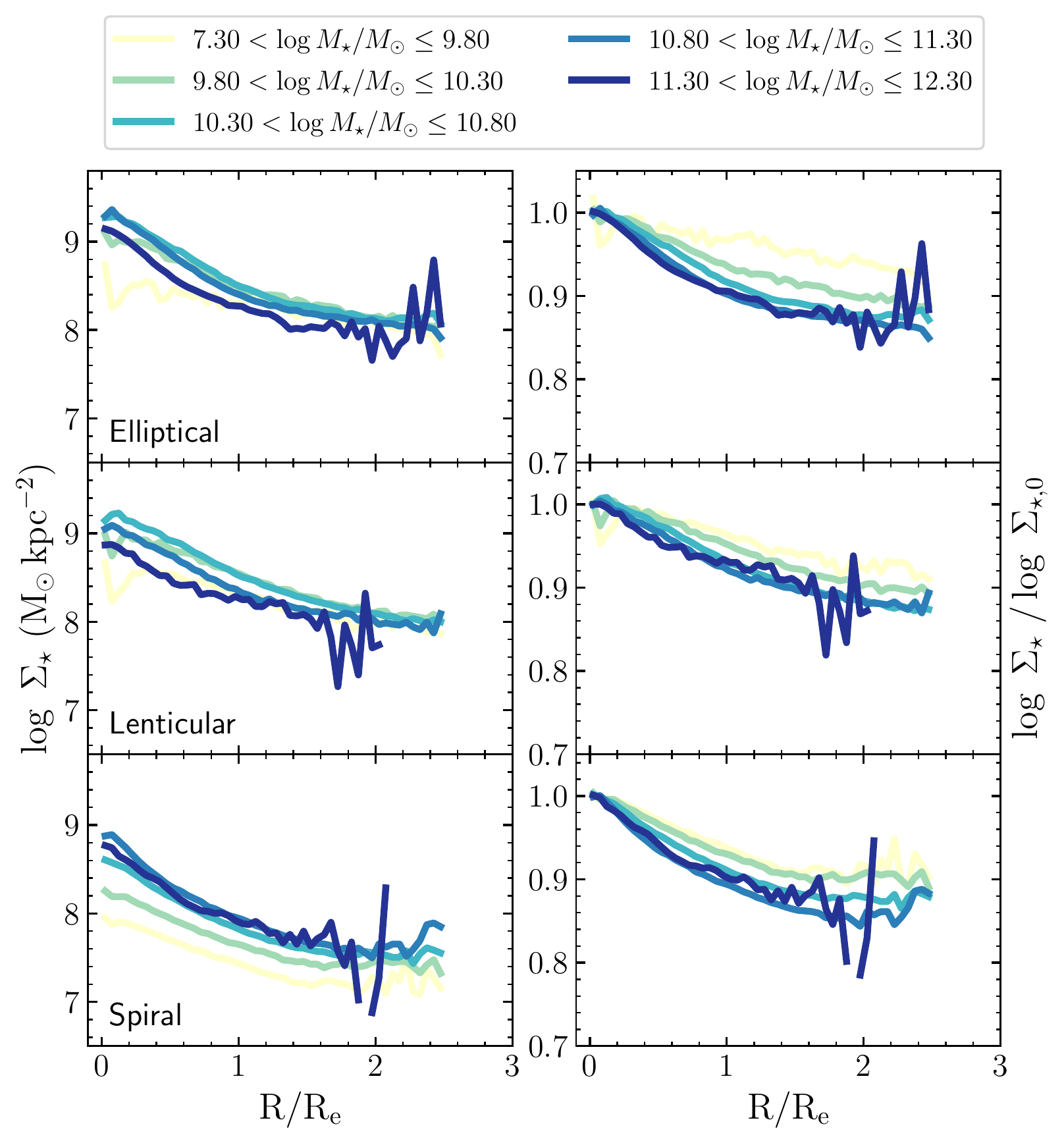}
    \caption{Radial profiles of stellar surface mass density in different bins of total stellar mass $M_\star$ (grey scale) and morphology (panels from top to bottom). Right hand panels show the equivalent profiles from the left hand side but normalised to the central surface mass density $\Sigma_{\star,0}$.}
    \label{fig:sbr}
\end{figure}

In the present work, we aim to relate these stellar population gradients with stellar mass surface density for which we present radial profiles in Fig. \ref{fig:sbr}. For simplicity we only show the medians for each mass-morphology bin. From this figure, it is apparent that $\Sigma_\star$ is monotonically decreasing with radius for all galaxies.  Higher mass galaxies have higher $\Sigma_\star$ at any given radius (except for the highest mass ETGs) and a more pronounced central upturn. Furthermore, spiral galaxies have lower $\Sigma_\star$ at fixed mass and radius than lenticulars or ellipticals. The trends are expected to change if the radius is not normalised by the effective light radius $R_\mathrm{e}$ due to the correlation between $R_\mathrm{e}$ and $M_\star$, as well as its dependence on morphology \citep[e.g.][]{vandeSande2019,Boardman2021}. These results are qualitatively consistent with previous studies \citep[e.g.][]{GonzalezDelgado2014a,GarciaBenito2017,Zibetti2020}.

The exclusively negative $\Sigma_\star$ gradients and the mostly negative metallicity gradients suggest that a local correlation should be found between $\Sigma_\star$ and $[Z/H]$ as local spatially resolved counterpart to the MZR, the $\rm r\Sigma_\star ZR$. At the same time, however, the differences seen in the gradients already point towards additional radially dependent drivers of stellar metallicity.

\subsection{Local stellar surface mass density-metallicity relation}

\begin{figure*}
	\centering
	\includegraphics[width=17cm]{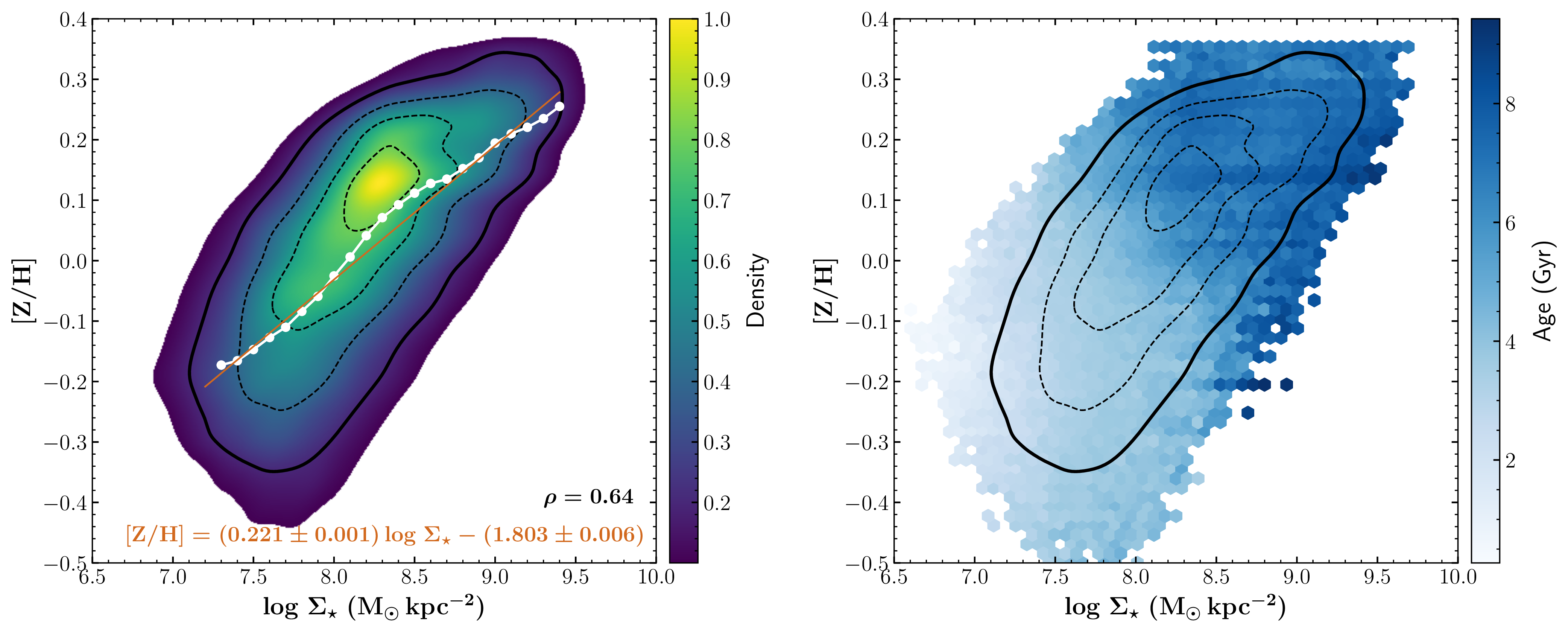}
    \caption{Spatially resolved stellar surface mass density-metallicity relation. \textit{Left:} density plot of all spatial bins from the complete sample out to maximal $2.5\, R_\mathrm{e}$. The data is smoothed with a Gaussian kernel density estimator. Contours enclose $20\%$, $40\%$, $60\%$ and $80\%$ of the sample. In white, we show median $[Z/H]$ in $\log \Sigma_\star$ bins of $0.1\,$dex width in the range $7.3< \log \Sigma_\star < 9.4$ and, in green, a linear regression to the data within the same range. The standard error on the median, calculated as $\sigma_\mathrm{err}=(\pi/2)\cdot (\sigma/\sqrt{N})$, where $\sigma$ is the standard deviation and $N$ the sample size per bin, is smaller than the marker size and thus not visible. The Spearman's rank correlation coefficient $\rho$ is shown in the bottom right. \textit{Right:} Same as left, but coloured according to the median age in each hexagonal bin without smoothing. Only bins with more than 20 data points are shown.}
    \label{fig:rMZR}
\end{figure*}

The local and spatially resolved stellar surface mass density-stellar metallicity relation, $\rm r\Sigma_\star ZR$\footnote{in the literature sometimes also called $\rm rMZR$, $\rm r\mu_\star ZR$ or $\Sigma_\star$-$[Z/H]$}, is shown in the left panel of Fig. \ref{fig:rMZR}. This is a density plot based on $> 2.6$ million Voronoi bins, radially rebinned as described above, from the complete working sample including galaxies from all morphologies, masses and from all radial positions out to $2.5\,R_\mathrm{e}$.

The figure shows a positive $\log \Sigma_\star$-$[Z/H]$ relation with a monotonic and close to linear increase of the binned median values. Due to a relatively broad distribution, we find a mild Spearman's rank correlation coefficient of $\rho=0.64$ with a p-value of $p \ll 0.001$. A linear regression yields the general equation:

\begin{equation}
[Z/H] = (0.221 \pm 0.001)\,\log \Sigma_\star - (1.803 \pm 0.006).
\end{equation}

In the right panel of Fig. \ref{fig:rMZR}, we display the same plot of the $\rm r\Sigma_\star ZR$ but colour-coded according to the median age. A clear positive trend of stellar age with both stellar metallicity and surface mass density is visible. However, the oldest and youngest population are not found at the highest and lowest metallicities but rather at the densest and less densest regions, respectively. Yet, at fixed mass density older populations tend to have higher metallicity.

\begin{figure}
	\includegraphics[width=\columnwidth]{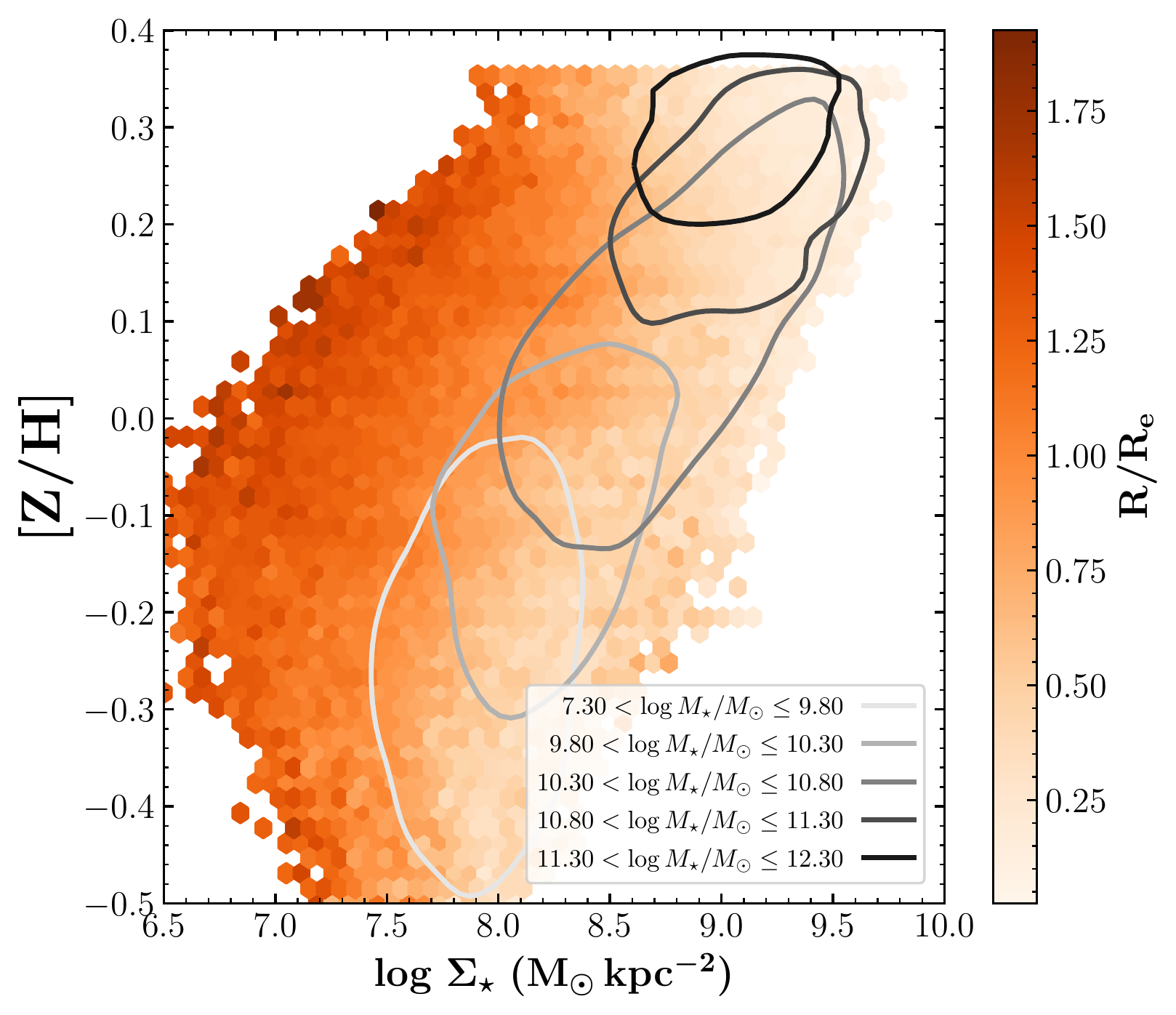}
    \caption{Same as Fig. \ref{fig:rMZR}, but coloured according to the median radius in each hexagonal bin without smoothing. Only bins with more than 20 data points are shown. Contours show smoothed $80\%$ density distributions from the inner part of galaxies, with $R \leq 0.25\,R_\mathrm{e}$, in different bins of total stellar mass $M_\star$.}
    \label{fig:rMZR_radius}
\end{figure}

From what we have seen, it seems clear that surface mass density drives stellar population properties locally, but we would also like to understand what causes the spread in the distribution and whether additional trends can be identified. A first step in this direction is made in Fig. \ref{fig:rMZR_radius}, where we present once again the $\rm r\Sigma_\star ZR$, this time coloured by the median galactocentric distance. 
At fixed metallicity, it is apparent that surface mass density is a decreasing function of radius corroborating the relations illustrated in Fig. \ref{fig:sbr}. Interestingly, at fixed $\Sigma_\star$, metallicity increases with radius. In fact, following same-coloured regions, stellar populations at the same median radius display a relatively tight $\rm r\Sigma_\star ZR$. This relation is shifted towards higher metallicities and lower surface mass densities at larger radii.

Furthermore, we explore how the total mass of the host galaxy affects the relation at fixed radius. On Fig. \ref{fig:rMZR_radius}, we superimpose contours for different mass bins and we consider only the central regions ($R\leq 0.25\,R_\mathrm{e}$). Each contour encloses 80\% of the subsample. The relation between $[Z/H]$ and $\Sigma_\star$ is reproduced at each mass bin (although it is not clear whether for the lowest mass bin we observe a very steep correlation or no correlation at all). As total mass increases, the distribution moves along a slightly curved $\rm r\Sigma_\star ZR$ with flattening slope towards higher masses. This behaviour is also representative for the trends at larger (but fixed) radii, which we not show here to maintain the clarity of the figure.

\begin{figure*}
	\centering
	\includegraphics[width=16cm]{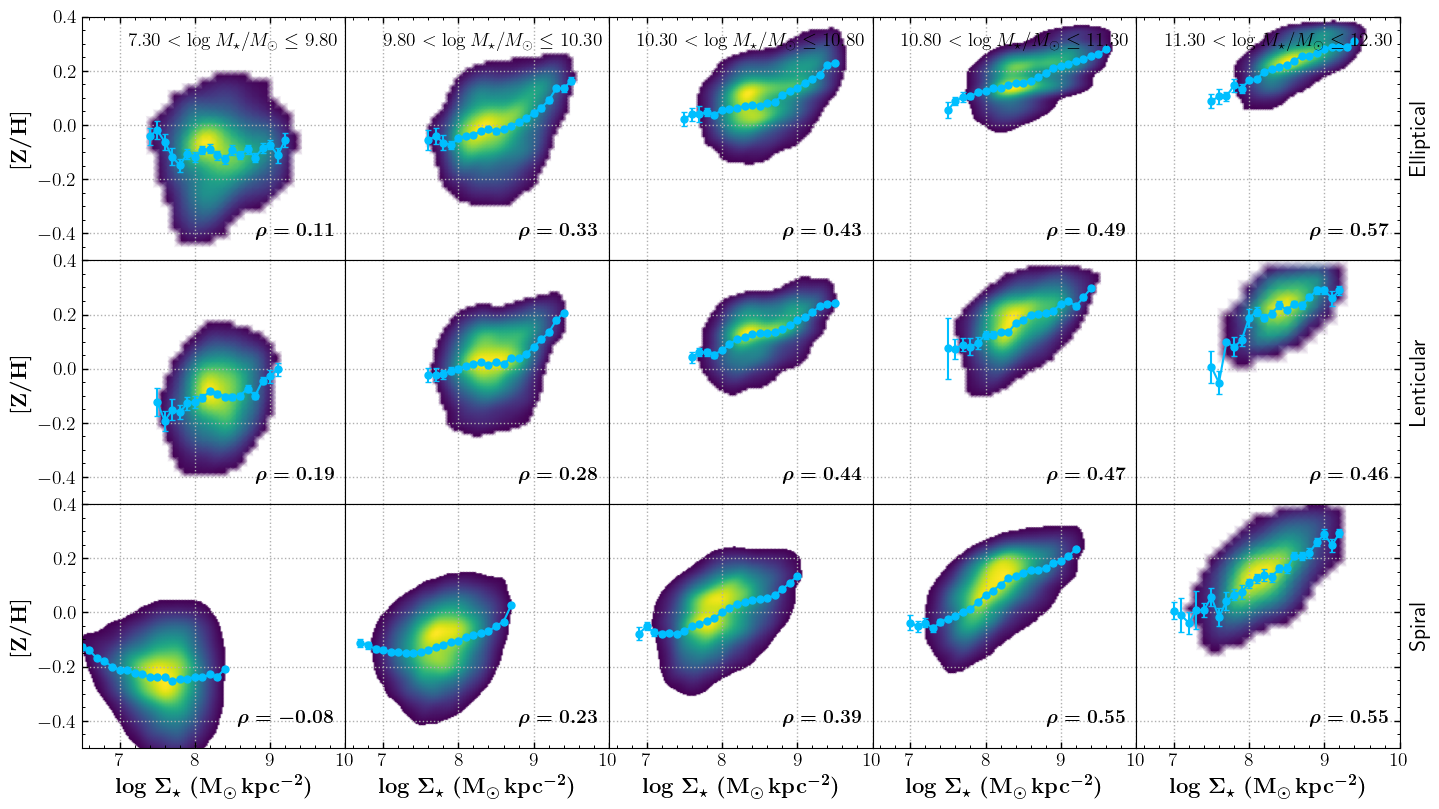}
    \caption{Spatially resolved stellar surface mass density-metallicity relation. Sames as left panel of Fig. \ref{fig:rMZR}, but here separated by total stellar mass and morphology of the respective galaxy. Sky blue dots show the median $[Z/H]$ in $\log\,\Sigma_\star$ bins of 0.1 dex.}
    \label{fig:rMZR_detail}
\end{figure*}

The dependence on total galaxy stellar mass and morphology looks different, if we do not separate the relation by radius but instead plot galaxy-wide $\rm r\Sigma_\star ZR$ separated by mass and morphology as shown in Fig. \ref{fig:rMZR_detail}. In contrast to the mass contours in Fig. \ref{fig:rMZR_radius}, this plot shows all spatial bins across each FoV; a mass-morphology separated version of Fig. \ref{fig:rMZR}.


A positive r$\Sigma_\star$ZR is found for all masses and morphologies except for the lowest mass bin. However, the correlation becomes weaker for lower galaxy masses and lower local mass densities. In fact, we observe an interesting bending in the relation between $\Sigma_\star \sim 10^8$ and $10^9\,\mathrm{M_\odot\,kpc^{-2}}$ at lower galaxy masses and almost independent of morphology. Similar upturns at high surface mass densities are also apparent in \citet[][Fig. 9]{Sanchez2020}. 

Thus, from Fig. \ref{fig:rMZR_detail} we conclude that the shape of the r$\Sigma_\star$ZR depends on total galaxy mass and morphology if radius is not fixed. This is qualitatively in good agreement with the results presented in \citetalias{GonzalezDelgado2014} and \citet{Sanchez2020}. We speculate that the disappearance of a mass density dependence at low total masses and low local mass densities might be an effect of increasing and outweighing importance of an additional radial-dependent driver of metallicity.


Altogether, the spatially resolved stellar population properties presented in Figs. \ref{fig:rMZR} - \ref{fig:rMZR_detail} paint a picture in which old and metal-rich populations are predominantly found in the centres of massive galaxies, the most metal-poor populations are found in the centres of the least massive galaxies, while the youngest populations are found in the outer parts of low-mass galaxies. The $\rm r\Sigma_\star ZR$ is reproduced at all masses except for the lowest mass bin. There is some indication that the relation is shifted towards higher metallicities at larger radii, but this cannot be clearly disentangled from an overlaying mass effect in Fig. \ref{fig:rMZR_radius}.

Very similar trends of age, $M_\star$ and $R$ in the $\Sigma_\star$-$[Z/H]$ plane are found in \citetalias{GonzalezDelgado2014} in an analsis of 300 CALIFA galaxies. However, the scatter in the $\rm r\Sigma_\star ZR$ is explained in \citetalias{GonzalezDelgado2014} by the impact of the global metallicity driver $M_\star$. While we come to the same conclusion that high metallicity samples come from massive galaxies, we point out that radius is an important parameter in that relation and that the scatter in the $\rm r\Sigma_\star ZR$ is significantly reduced if the relation is shown at fixed radius. The flattening of the relation in the inner regions of high mass galaxies is in very good agreement with \citetalias{GonzalezDelgado2014} and is interpreted by these authors as due to a dominant role of $M_\star$ as driver of metallicity in spheroids as opposed to a $\Sigma_\star$ dominated regime in discs. Contrary to that interpretation, \citet{Zibetti2020} show that metallicity in massive ETGs is primarily driven by surface mass density and only secondarily modulated by galaxy mass. This is attributed by the authors to steeper metallicity gradients in their work. When we compare the $\rm r\Sigma_\star ZR$ in Fig. \ref{fig:rMZR} to the median relation for ETGs in \citet{Zibetti2020}, we find a remarkably good agreement in the overlapping range of $8< \log \Sigma_\star < 9.4$ with metallicity increasing from slightly sub-solar to $[Z/H]\sim 0.22$. However, if we consider only ETGs with $\log\,M_\star/M_\odot>10.3$ in Fig. \ref{fig:rMZR_detail}, our values are shifted towards slightly higher metallicities.

In \citet{Zhuang2019}, the authors report a tight local relation between dynamical surface mass density and stellar metallicity within $1\,R_\mathrm{e}$ from measurements of 244 CALIFA galaxies across all morphological types. This dynamical relation is consistent with our $\rm r\Sigma_\star ZR$, in particular for the central regions, and strengthen further our results.

\subsection{Metallicity trends at fixed surface mass density}

\begin{figure*}
	\centering
	\includegraphics[width=17cm]{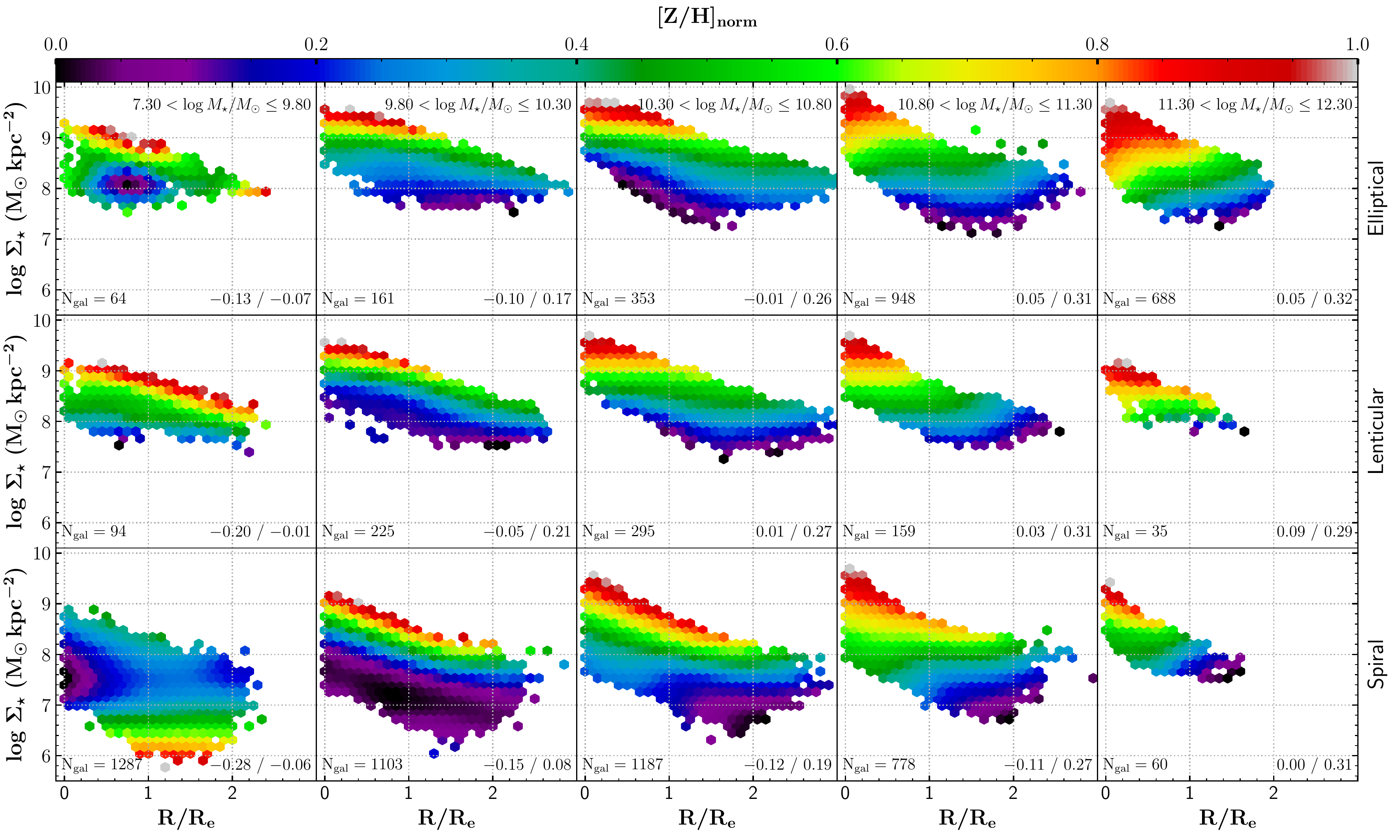}
    \caption{Stellar surface mass density versus radius coloured by median stellar metallicity for individual spatial bins. The grid shows the global mass-morphology plane of the host galaxies. The number of galaxies that contributed to each panel are shown in the lower left corner. We use individual colour bar limits in each panel to highlight subtle trends. The minimum and maximum of each colour bar is shown in the lower right corner of the corresponding panel. The data is smoothed using the LOESS method as described in the main body of the text.}
    \label{fig:MRR}
\end{figure*}

In Fig. \ref{fig:MRR}, we present the local relations between $\Sigma_\star$, $[Z/H]$ and $R/R_\mathrm{e}$ from a different perspective. In this plot, we focus on how the metallicity changes as a function of radius at fixed surface mass density in the total stellar mass-morphology plane. We thereby test in more detail the hypothesis drawn from Fig. \ref{fig:rMZR_radius} that metallicity increases with radius at fixed mass density and fixed total mass. We plot again single spatial bins from the whole population of galaxies divided by mass and morphology. The colour of each hexagonal area shows the median $[Z/H]$ of all contributing spatial bins in the corresponding $\Sigma_\star$-$R/R_\mathrm{e}$ intervals.

We employ Locally Weighted Regression \citep[LOESS;][]{Cleveland1988} in the python implementation from \citet{Cappellari2013} to smooth the data in two dimensions. This method has been successfully tested in several works to study and recover mean trends of stellar population properties in galaxies \citep[e.g.][]{McDermid2015,RosadoBelza2020,Boardman2021}.

First of all, at almost any radial location in any part of the grid, we clearly see increasing metallicity with increasing surface mass density, i.e. a r$\Sigma_\star$ZR is found at any galactocentric distance. Additionally and importantly, if we look at these plots at fixed $\Sigma_\star$ instead of at fixed radius, we observe that for most of the mass densities, galaxy masses and morphologies $[Z/H]$ is increasing with radius. The radial dependence is clearest for spiral galaxies and in low mass ETGs ($\log M/M_\odot < 10.8$) except for the lowest surface mass densities. It disappears in higher mass ETGs.

\begin{figure}
	\centering
	\includegraphics[width=\columnwidth]{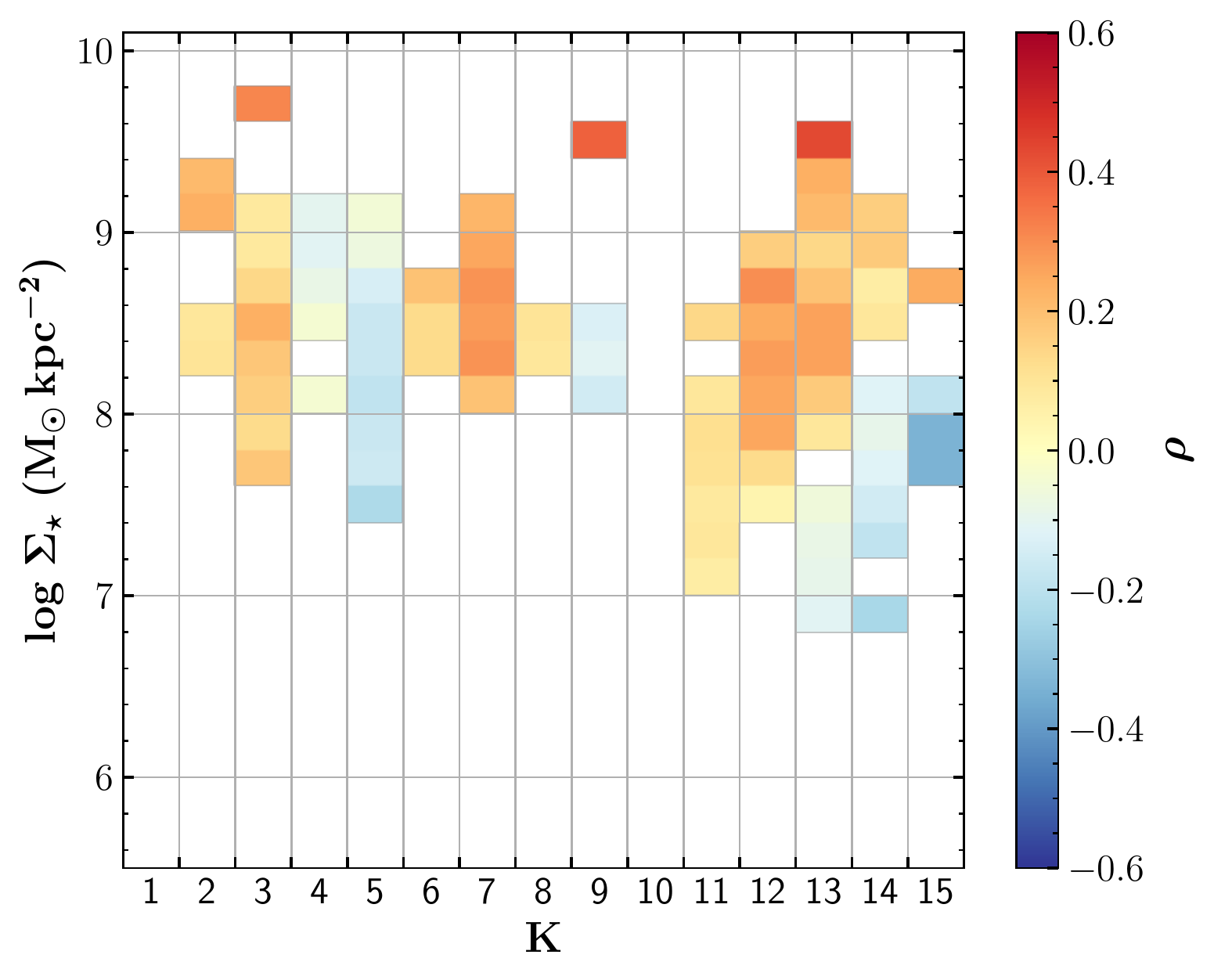}
    \caption{Spearman's rank correlation coefficients $\rho$ testing a correlation between $[Z/H]$ and $R/R_\mathrm{e}$ in fixed bins of $\Sigma_\star$. On the x-axis, $K$ is a simple numbering of the 15 panels in the mass-morphology grid with $K$=1..5 for ellipticals with increasing mass, $K$=6..10 lenticulars and $K$=11..15 spirals. $\rho$ is only shown if the $p$-value p<0.05 and if the number of datapoints is larger than 20.}
    \label{fig:stats}
\end{figure}

We explore the strength and significance of this radial dependence further in Fig. \ref{fig:stats}. For each panel across the mass-morphology grid, we compute the Spearman's rank correlation coefficient between $[Z/H]$ and $R/R_\mathrm{e}$ in fixed $0.2\,\mathrm{dex}$ bins of $\log \Sigma_\star$. We show only statistically significant correlations by limiting the corresponding $p$-value to $p<0.05$. Positive but weak correlations are found for $K$=2,3,6,7,8,11,12,13 and at high mass densities in $K$=14,15. This corresponds to all galaxies below $\log M_\star/M_\odot < 10.8$ except for the lowest-mass ellipticals, as well as to high-density regions in massive spirals. The dependence is slightly negative in high-mass ETGs and at low-density regions in high-mass spirals. These statistics corroborate the average trends observed in Fig. \ref{fig:MRR}.





In addition to local trends, we can also confirm in Fig. \ref{fig:MRR} that metallicity is globally driven by galaxy mass and morphology, in that $[Z/H]$ is larger for more massive galaxies (from left to right) and for earlier types (from bottom to top) even at fixed $\Sigma_\star$ and galactocentric distance.  However, the cross comparison is admittedly not an easy exercise to do in this representation given the individual colour bar limits and it is better represented in the previous sections. 

From the representation of trends of stellar metallicity in Fig. \ref{fig:MRR}, it is clear that negative metallicity gradients in local galaxies are largely imposed by radially monotonically decreasing surface mass density and, in fact, metallicity is not decreasing `steep enough' from what mass densities would drive. At fixed surface mass density metallicity is increasing with galactocentric distance. This is true for most low- to intermediate-mass galaxies and at high $\Sigma_\star$ in massive spirals. This result calls for additional driver of stellar metallicity that enhances metallicity more strongly at larger radii or/and dilutes it at smaller radii.

\section{Discussion}
\label{sect:discuss}

The observed metallicity in stars is a result of the chemical evolution and star formation history. To increase the metallicity, the ejected gas from a population of stars needs to be recycled into a new stellar generation. The more generations of stars are formed the higher the possibility to build-up metallicity, but more importantly, the pre-enriched gas needs to be efficiently recycled. The relation between mass and metallicity is usually understood as a consequence of the ability to retain enriched gas ejecta. The higher the mass, the higher is the potential well that keeps gas back and efficiently recycles it to new stars \citep[e.g.][and references therein]{Thomas1999}.

Reality is, however, more complex than this closed-box scenario and cosmological simulations have shown the importance of inflows, outflows, mergers and radial migration in shaping the radial distribution of stellar populations in galaxies \citep[e.g.][]{Sellwood2002,Roskar2008,Minchev2010,ElBadry2016,Collacchioni2020,Grand2019}.

Inverted or positive metallicity gradients at higher redshifts have been explained by a dilution of the ISM in the centre due to pristine metal-poor gas inflow in the central regions \citep{Cresci2010} and, additionally, outflows of metal-rich gas from the centre \citep{Troncoso2014}. In the galactic fountain scenario this metal-rich gas is re-accreted onto the galaxy out to large radii \citep[e.g.][]{Oppenheimer2010}. Higher mass galaxies are able to retain outflows earlier in cosmic history and thereby enable a rapid metal enrichment \citep{Tremonti2004,Muratov2017}. Furthermore, central metallicity dilution due to inflows can also be merger-induced as shown from cosmological galaxy formation simulations in \citet{Bustamante2018}, thus, environment is likely to impact the chemical evolution.

A number of cosmological simulations have shown that these and similar processes can shape the observed gas and stellar metallicity gradients in local galaxies. From the zoom-in Auriga simulations, \citet{Grand2019} reports that most of the material in redshift zero stars has been ejected and reaccreted in form of galactic fountains at least once \citep[cf.][]{Brook2014,Uebler2014,Christensen2016}. By comparing different metal loading factors, they show that galactic fountains play an essential role in efficiently redistributing metal-enriched gas and thereby in flattening of the stellar metallicity gradient. However, the fractional contribution of different accretion channels is heavily debated in the literature. For example, \citet{Mitchell2020} find that first infall accretion clearly dominate recycled inflow in the EAGLE simulations, while \citet{AnglesAlcazar2017} report fractions for the FIRE simulations in between that of Auriga and EAGLE.

At later times in cosmic history metal-rich gas accretion, e.g. from intergalactic transfer \citep{AnglesAlcazar2017}, onto the outer parts of galaxies, triggering or maintaining star formation in the outer disc, is also contributing to increased metallicity at large galactocentric distances.

In addition to gas flows, stellar migration presents another mechanism that has been shown to flatten stellar metallicity gradients both in $N$-body simulations as well as in large cosmological simulations \citep[e.g.][]{Roskar2008,DiMatteo2013,Grand2015,ElBadry2016,Buck2020,Vincenzo2020}. Nevertheless, the extent to which stellar migration happen in real galaxies has proven to be difficult to assess in observations so far.

Finally, inside-out quenching -- either due to morphological quenching \citep{Martig2009}, central compaction \citep{Tacchella2016} or feedback from active galactic nuclei \citep{GuoK2019} -- constitutes another effective way of halting the chemical evolution in the centres of galaxies first while outer parts continue to increase their metal content \citep[e.g.][]{Lacerna2020}. 

In summary, there are a variety of processes seen in simulations that affect the chemical evolution of galaxies and are able to substantially alter the radial distribution of metals. Our observations presented in this paper show clear evidence for a primarily mass density-driven metal distribution. This distribution, nevertheless, requires additionally a process (or conglomeration of processes) that promotes enhanced chemical enrichment at larger galactic radii.

In order to allow for a better connection of the physical processes in simulations with observations and to quantitatively compare the predicted and observed distributions of stellar populations in galaxies, it will be interesting to explore whether this observed $\rm r\Sigma_\star ZR$ can be reproduced in mock observations of cosmological simulations.








\section{Conclusions}
\label{sect:conclusion}

This work constitutes a detailed analysis of spatially resolved stellar populations using a very large sample of 7439 galaxies from the MaNGA survey. The sample size allows us to split galaxies by mass and morphology and to push simultaneously the statistical uncertainties to very low values. We focus specifically on the local variations of stellar metallicity with stellar surface mass density and galactocentric distance in the global mass-morphology plane. Our results can be summarised as follows:

\begin{itemize}
\item We successfully reproduce the global age-mass and mass-metallicity relations for galaxies of different morphologies.
\item We present radial age, metallicity and surface mass density distributions and find that metallicity gradients are mostly negative but vary strongly with mass, while age profiles are mostly mildly positive in ETGs and negative but shallow in LTGs. Surface mass density profiles are consistently negative with steeper gradients for more massive galaxies.
\item We explore in detail the spatially resolved surface mass density-metallicity relation $\rm r\Sigma_\star ZR$ and find that mass density drives stellar metallicity locally, which connects negative metallicity gradients -- as found in most galaxies -- to negative surface mass density gradients. In addition, the $\rm r\Sigma_\star ZR$ is modulated by radial distance.
\item In particular, we find that at fixed mass density, metallicity increases with radius. This holds true for the largest part of the sample below $\log M/M_\odot \sim 10.8$ except for the lowest-mass ellipticals. For $\log M/M_\odot > 10.8$ the positive trend is only seen in spirals at high surface mass densities, while at low $\Sigma_\star$ the correlation is slightly negative. The positive radial dependence is strongest for spiral galaxies. 
\item The previous result requires an additional driver of stellar metallicity that promotes chemical enrichment in the outer parts of galaxies more strongly than in the inner parts. We discuss gas accretion, outflows, quenching and stellar migration as possibilities. A more direct comparison of the effect of these processes as seen in cosmological simulations on the distribution of stellar populations as seen in the observation will help to shed further light on that matter.
\end{itemize}

\section*{Acknowledgements}

We thank the anonymous referee for his comments that helped to improve the paper. The Science, Technology and Facilities Council is acknowledged for support through the Consolidated Grant Cosmology and Astrophysics at Portsmouth, ST/S000550/1. JL is supported by the National Science Foundation under Grant No. 2009993. J B-B acknowledges support from the grant IA-100420 (DGAPA-PAPIIT, UNAM), and funding from the CONACYT grants CF 19-39578, CB-285080 and FC-2016-01-1916.

Numerical computations were done on the Sciama High Performance Compute (HPC) cluster which is supported by the ICG, SEPnet and the University of Portsmouth.

Funding for the Sloan Digital Sky Survey IV has been provided by the Alfred P. Sloan Foundation, the U.S. Department of Energy Office of Science, and the Participating Institutions. SDSS acknowledges support and resources from the Center for High-Performance Computing at the University of Utah. The SDSS web site is \url{www.sdss.org}.

SDSS is managed by the Astrophysical Research Consortium for the Participating Institutions of the SDSS Collaboration including the Brazilian Participation Group, the Carnegie Institution for Science, Carnegie Mellon University, the Chilean Participation Group, the French Participation Group, Harvard-Smithsonian Centre for Astrophysics, Instituto de Astrof\'isica de Canarias, The Johns Hopkins University, Kavli Institute for the Physics and Mathematics of the Universe (IPMU) / University of Tokyo, the Korean Participation Group, Lawrence Berkeley National Laboratory, Leibniz Institut f\"ur Astrophysik Potsdam (AIP), Max-Planck-Institut f\"ur Astronomie (MPIA Heidelberg), Max-Planck-Institut f\"ur Astrophysik (MPA Garching), Max-Planck-Institut f\"ur Extraterrestrische Physik (MPE), National Astronomical Observatories of China, New Mexico State University, New York University, University of Notre Dame, Observatório Nacional / MCTI, The Ohio State University, Pennsylvania State University, Shanghai Astronomical Observatory, United Kingdom Participation Group, Universidad Nacional Aut\'onoma de M\'exico, University of Arizona, University of Colorado Boulder, University of Oxford, University of Portsmouth, University of Utah, University of Virginia, University of Washington, University of Wisconsin, Vanderbilt University, and Yale University.

\section*{Data Availability}

We shall release the data underlying this article only through the
official releases of the SDSS-IV.




\bibliographystyle{mnras}
\bibliography{NewDatabase.bib} 



\appendix

\section{Example 2D maps from the MaNGA Firefly VAC}

\begin{figure*}
	\centering
	\includegraphics[width=16cm]{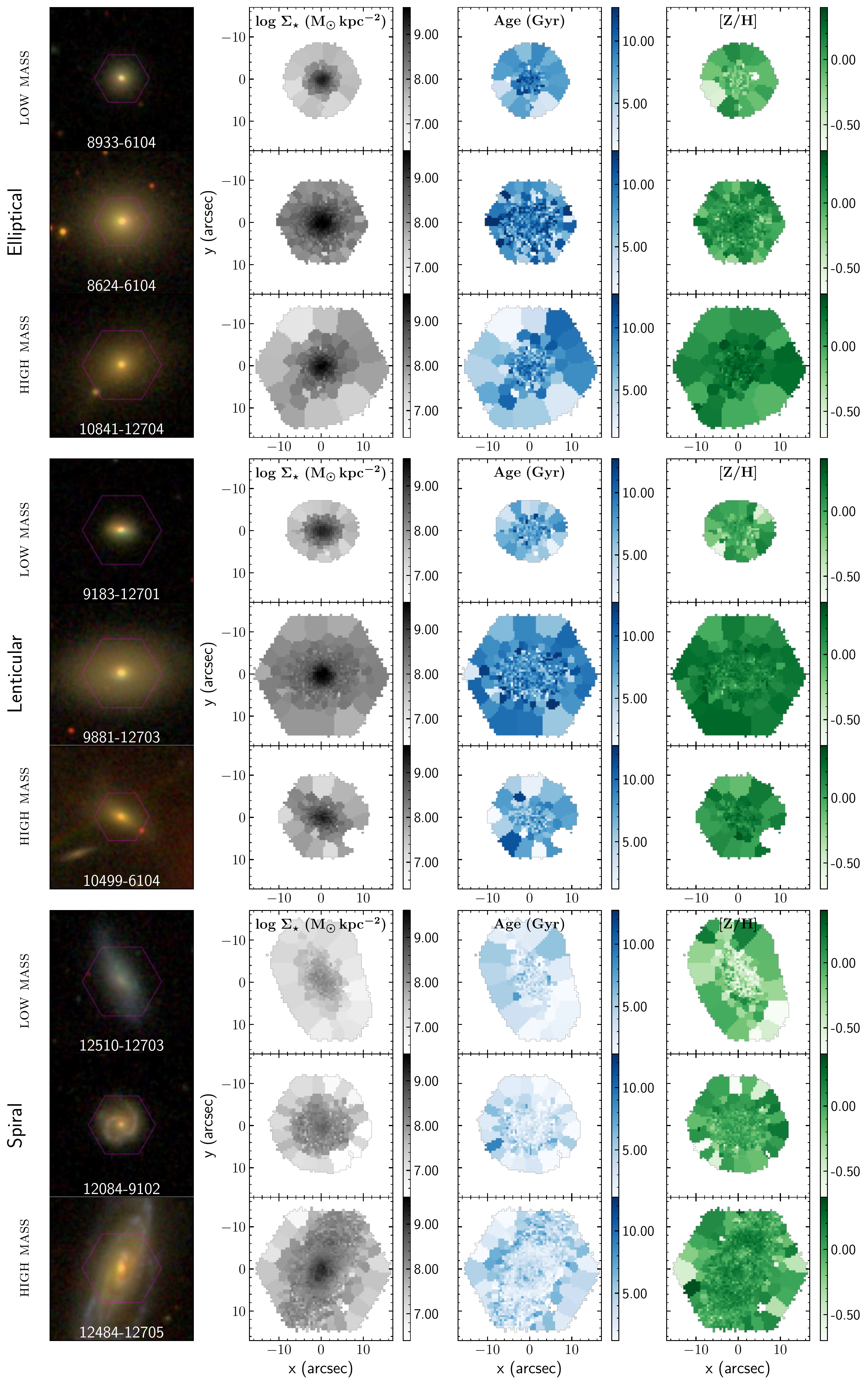}
    \caption{Example 2D maps from the MaNGA Firefly VAC.}
    \label{fig:galEx}
\end{figure*}

In Fig. \ref{fig:galEx} we show example maps of 2D stellar population properties from the MaNGA Firefly VAC.

\bsp	
\label{lastpage}
\end{document}